\documentclass[twocolumn]{aastex61}

\shortauthors{Demirci et al.}
\begin{document}

\title{Is There a Temperature Limit in Planet Formation at 1000\,K?}

\correspondingauthor{Tunahan Demirci}
\email{tunahan.demirci@uni-due.de}

\author{Tunahan Demirci}
\affil{Faculty of Physics, University of Duisburg-Essen, Lotharstr. 1, D-47057 Duisburg, Germany}

\author{Jens Teiser}
\affiliation{Faculty of Physics, University of Duisburg-Essen, Lotharstr. 1, D-47057 Duisburg, Germany}

\author{Tobias Steinpilz}
\affiliation{Faculty of Physics, University of Duisburg-Essen, Lotharstr. 1, D-47057 Duisburg, Germany}

\author{Joachim Landers}
\affiliation{Faculty of Physics, University of Duisburg-Essen, Lotharstr. 1, D-47057 Duisburg, Germany}
\affiliation{Center for Nanointegration Duisburg-Essen (CENIDE), University of Duisburg-Essen, Carl-Benz-Str. 199, D-47057 Duisburg, Germany}

\author{Soma Salamon}
\affiliation{Faculty of Physics, University of Duisburg-Essen, Lotharstr. 1, D-47057 Duisburg, Germany}
\affiliation{Center for Nanointegration Duisburg-Essen (CENIDE), University of Duisburg-Essen, Carl-Benz-Str. 199, D-47057 Duisburg, Germany}

\author{Heiko Wende}
\affiliation{Faculty of Physics, University of Duisburg-Essen, Lotharstr. 1, D-47057 Duisburg, Germany}
\affiliation{Center for Nanointegration Duisburg-Essen (CENIDE), University of Duisburg-Essen, Carl-Benz-Str. 199, D-47057 Duisburg, Germany}

\author{Gerhard Wurm}
\affiliation{Faculty of Physics, University of Duisburg-Essen, Lotharstr. 1, D-47057 Duisburg, Germany}



\begin{abstract}

Dust drifting inward in protoplanetary disks is subject to increasing temperatures. In laboratory experiments, we tempered basaltic dust between 873\,K and 1273\,K and find that the dust grains change in size and composition. These modifications influence the outcome of self-consistent low speed aggregation experiments showing a transition temperature of 1000\,K. Dust tempered at lower temperatures grows to a maximum aggregate size of $2.02 \pm 0.06$\,mm, which is $1.49 \pm 0.08$ times the value for dust tempered at higher temperatures. A similar size ratio of $1.75 \pm 0.16$ results for a different set of collision velocities.
This transition temperature is in agreement with orbit temperatures deduced for observed extrasolar planets. Most terrestrial planets are observed at positions equivalent to less than 1000\,K.
Dust aggregation on the millimeter-scale at elevated temperatures might therefore be a key factor for terrestrial planet formation.

\end{abstract}




\section{Introduction}

\begin{figure}[h]
	\includegraphics[width=\columnwidth]{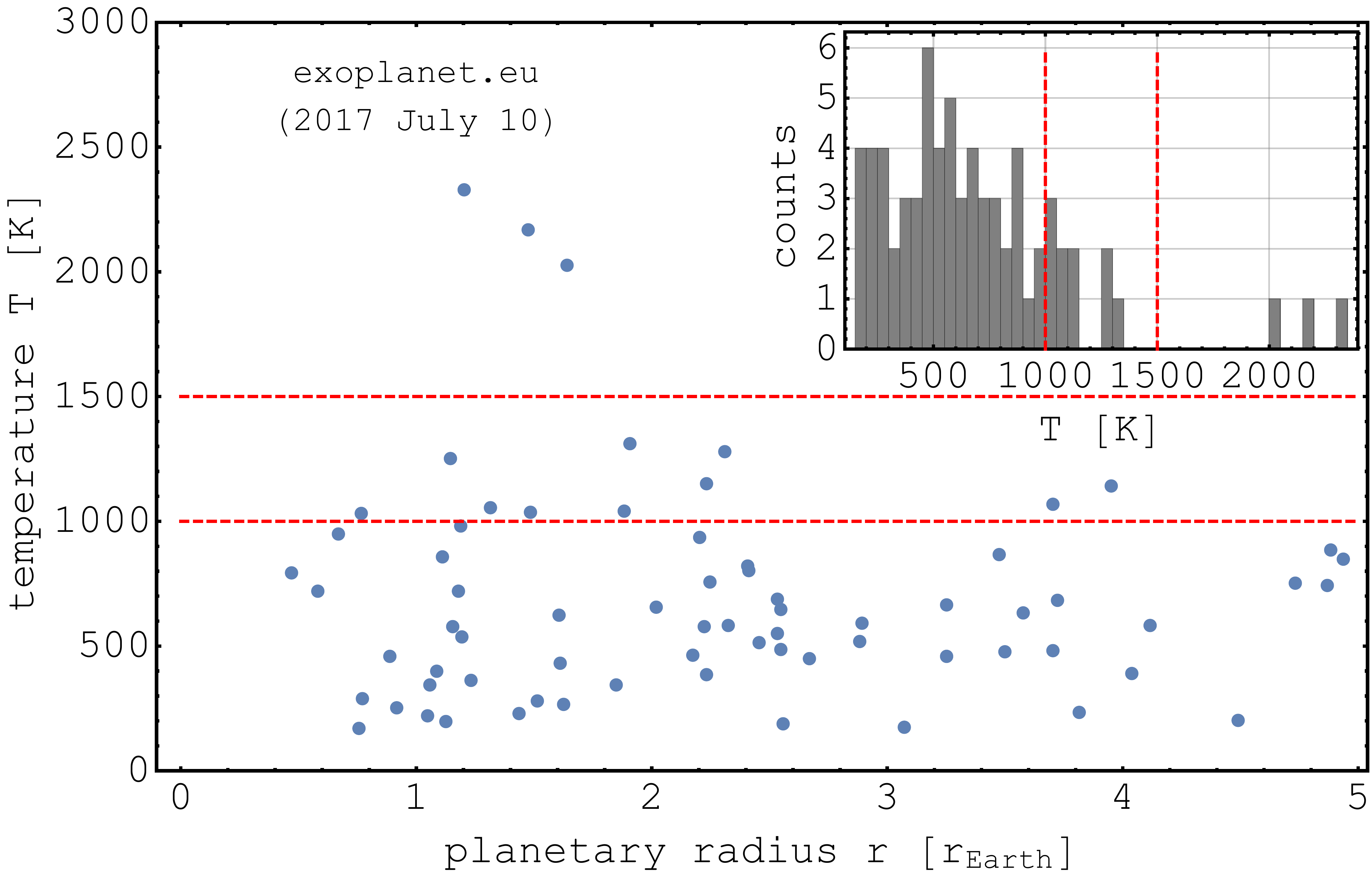}
	\caption{\label{fig.exos}Temperatures of observed planets with radii smaller than 5
		$r_{\mathrm{Earth}}$. Data are taken from \cite{Exoplanet.eu}. The 1000\,K limit as well as the upper limit of the condensation region at 1500\,K are highlighted. The inset summarizes the distribution of all shown planets over temperature.}
\end{figure}

The number of observed extrasolar planets increases continuously \citep[www.exoplanet.eu]{Schneider2012}. This now allows a statistical view on the distribution of certain parameters. The semi major axes of gas planets can be rather small, placing them in close proximity to their host stars (hot Jupiters). This is evidence of migration as giants are supposed to be formed beyond the water snowline \citep{Kley2000,Nelson2003,Alibert2005,Benz2014,Duermann2015,Duermann2017}. For smaller Earth-like planets
the situation is different. Although detection by transits and radial velocity works best for planets with small orbital periods, significantly fewer terrestrial planets are found in the hot environment close to the star compared to the cooler outer parts \citep{Exoplanet.eu}. This is shown in Figure \ref{fig.exos} with data taken from \citet{Exoplanet.eu}.
However, we do not plot the radial distance here, instead using the temperature attributed to the planet. For a small planet this should essentially be the temperature of radiative equilibrium with the stellar radiation. 

Overlaid are two lines of constant temperature. The 1000\,K line corresponds to a change in aggregation observed in our work (see below). 1500\,K is approximately the upper limit of the condensation region for typical silicates. The database is still one with low statistics, and detailed dependencies on temperature might be discussed. There is, e.g., a maximum at about 500\,K, but the distribution might just as well be considered flat in a wide range. None of these statements are statistically significant at the moment. Nevertheless, there is a clear gap in the occurrence of planets between about 1000\,K and 1500\,K.

For a planet forming at a larger distance from the star, the inward drift would not need to stop at the 1000\,K line. Also, from a more general perspective of late planet formation phases there is no obvious reason why planets should not form, e.g. at 1100\,K as likely as at 900\,K. That leaves the question whether this temperature gap is inherited from the early phases of planet formation by aggregation at elevated temperatures.

More often than not, planet formation and its limits are considered in light of radial distance. It might be argued that the lack of planets at higher temperatures might be due to a closer distance and a related lack of material for planet formation. Certainly, temperature and distance are related, but we note that they are not the same. Small changes in stellar mass already lead to large luminosity and significant temperature variations. However, the local material available will not significantly change. Also, it is currently often argued that inner planetary systems are formed from 
material drifting inward \citep{Boley2013,Wurm2013,Hu2014}.
Inward drift is a classical mechanism for redistribution of matter \citep{Weidenschilling1977}. As a special aspect, e.g., \cite{Ebel2011} suggested that Mercury might have formed from dusty matter (chondritic porous interplanetary particles) drifting inward that were reprocessed in a region of higher temperatures with a composition that is then very different from solar composition. Considering such work, one motivation for our experiments is a radial inward drift of a mix of materials (see below). The mass of material available for planet formation does not necessarily depend only on the radial distance to the star. We also note that current planet temperatures and temperatures at the time of dust aggregation might not be the same. They might be comparable if the relevant aggregation occurred in a passive disk. They might have been higher in a still actively accreting disk. However, accretion of dust at these active times might have removed these aggregates again. In any case, we do not claim to have the only answer to planet formation in the inner disk at high temperature, but we think it is curious that there is a temperature limit in the exoplanet database that might be reason enough to provide this somewhat different point of view here and take it as motivation for our experimental work.

Collisional growth of micrometer dust to millimeter aggregates is a first step in the process of planet formation. Numerous experimental \citep{Blum2008,Guettler2010,Kruss2017} and theoretical \citep{Dominik1997,Wada2011} studies showed that sticking collisions easily provide porous dust aggregates in the millimeter range. Growth saturates at this size though, as bouncing between particles becomes dominant and further growth is stalled. This is therefore also known as the bouncing barrier \citep{Zsom2010,Wada2011,Drazkowska2014,Kelling2014,Kruss2016, Kruss2017}. Its existence is a robust finding. Only the final size of the bouncing agglomerates might be a "free" parameter depending on the initial conditions as, e.g. the grain size distribution, disk properties, collision velocities, or material. 

In a number of current planet formation models, gravity has to take over from here  \citep{Chiang2010}. This requires mechanisms to concentrate the bouncing aggregates  \citep{Klahr1997,Johansen2006,Chiang2010}. These are most efficient if dust aggregates are larger than a certain minimum size (minimum Stokes number). Depending on the individual model, the critical particle size varies from decimeter down to millimeters. Recent findings set lower limits to the minimum Stokes number than before \citep{Yang2017}. Larger dust agglomerates are favorable for all of these models, as they can be concentrated more easily. However, it should be mentioned that there is also a maximum Stokes number above which the streaming instability does not seem to work. In a simplified view, there are two important sizes at the transition from collisional growth and concentration mechanisms.
The maximum size of particles grown at the bouncing barrier and the minimum size needed, e.g. for streaming instabilities do not largely overlap. 
In fact, it might rather be a problem that they do not overlap at all \citep{Drazkowska2014}. It is therefore not a trivial detail but instead is crucial how large aggregates can grow via coagulation and even a small difference might decide whether planetesimals locally form or not. 

Considering a rather crude distinction in composition and temperature, it is well known that the material has a direct influence on collisional growth.  Water ice beyond the
(water) snowline, for example, is considered to be more sticky than silicates \citep{Dominik1997,Gundlach2011,Okuzumi2012,Aumatell2013,Gundlach2015,Musiolik2}.
Non-polar $\rm CO_2$ ice on the other side behaves like silicates beyond its snowline \citep{Musiolik1}. But these are only general trends.  In this scheme, silicates should have one behavior everywhere beyond their "snowline", which is more commonly referred to as condensation region at about $1350-1500$\,K \citep{Scott2007}. This might be questionable though. In particular for a mixture of refractory grains, the actual temperature even far below the sublimation might be important. At high but not extreme temperatures re-crystallization and (partial) sintering might already change the mechanical properties of agglomerates as well as the collision behavior of single grains. 
\cite{deBeule2017} considered changes in (static) sticking properties for tempered dust samples for the first time.
Here, we study the influence of the tempering temperature on the bouncing limit of dust aggregates consisting of a mixture of minerals (olivines, pyroxenes, iron oxides, ...).

\section{Growth experiments}

\begin{figure}[h]
	\includegraphics[width=\columnwidth]{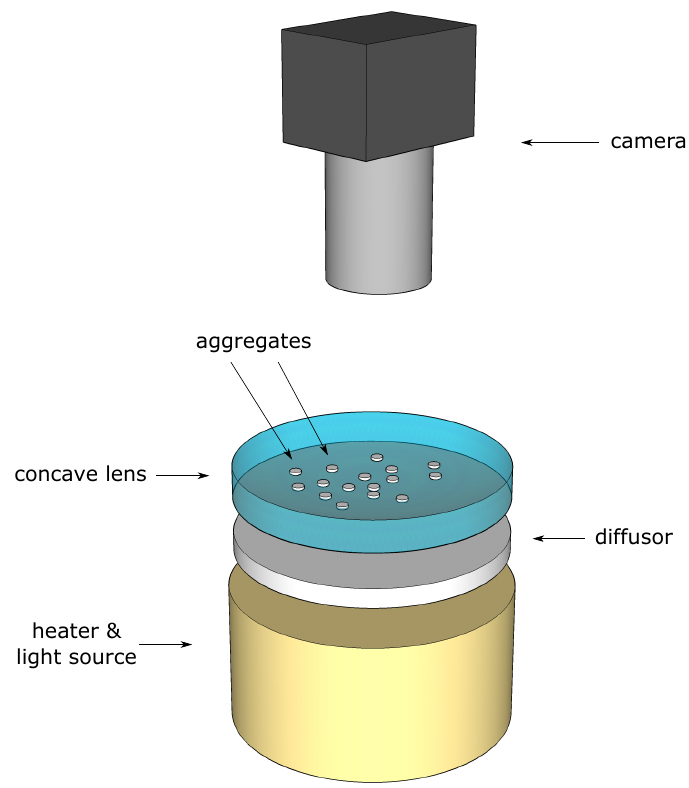}
	\caption{\label{fig.setup}Laboratory levitation setup to study the aggregation
		of dust at low velocities \citep{Kruss2016,Kruss2017}.}
\end{figure}

The basic experimental setup underlying this work  has been used in a number of earlier studies on dust aggregation \citep{Jankowski2012, Kelling2014, Kruss2016, Kruss2017} (Figure \ref{fig.setup}). 
The principle way it works is as follows.
In a low-pressure environment (here, $p \approx 500$\,Pa) capillaries on the order of the mean free path of the gas molecules and heated on one end efficiently pump gas from the cold to the warm side \citep[thermal creep]{Knudsen1909,Muntz2002}. 
A porous dust aggregate acts like a collection of capillaries \citep{deBeule2015,Koester2017,Steinpilz2017}. Therefore, if dust aggregates are placed on a hot surface at low ambient pressure, gas flows from the cold top through the aggregate toward its bottom. 
This leads to an overpressure \citep{Knudsen1909}. If the overpressure is large enough the dust aggregates are levitated on the air cushion they produce. A large number of levitated aggregates can be generated this way that then collide with velocities in the mm\,s$^{-1}$ to cm\,s$^{-1}$ range.

This velocity is in the range expected for dust in protoplanetary disks at the bouncing barrier, where aggregates no longer grow by hit-and-stick but preferentially bounce off each other \citep{Guettler2010,Zsom2010}. 
\cite{Kruss2017} showed that a bouncing barrier evolves self-consistently if small ($\lesssim 100\,\mu$m) dust particles are placed on the heater.
How large the dust grows at maximum has not been studied with this setup before.
For a given experimental setting (pressure $p$, temperature $T$) it will depend on the sticking properties of the dust grains. 

\begin{figure}[h]
	\includegraphics[width=\columnwidth]{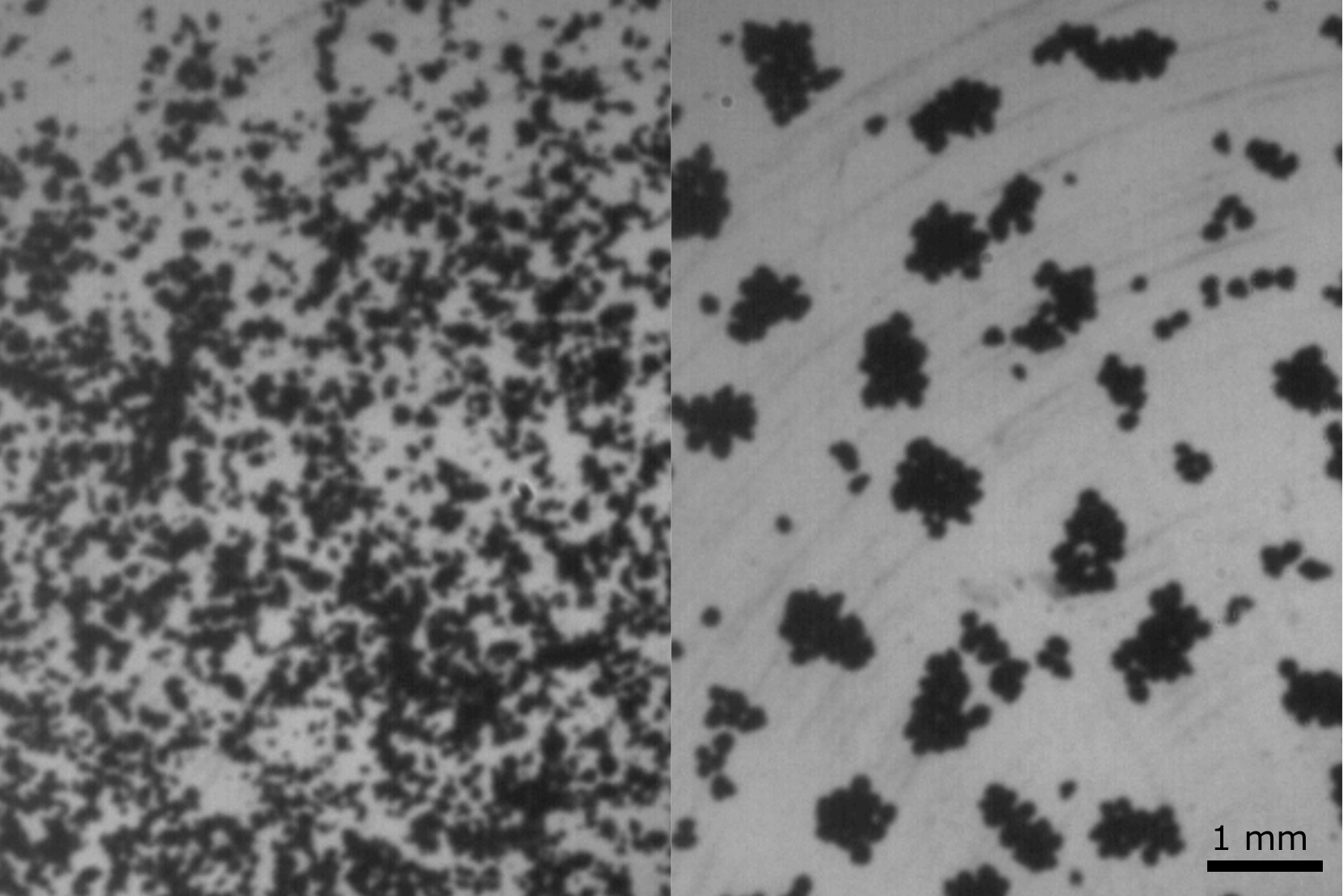}
	\caption{\label{fig.examples}Snapshot of the initial dust placed on the heater and the final size of aggregates after 15 minutes of growth. The initial sample was sieved onto the heater through a $125\, \rm \mu m$ mesh.}
\end{figure}

We use this setup here to test whether a dust sample that is subject to tempering for some time leads to a different final outcome. 
A snapshot of a final aggregate distribution after 15 minutes is shown in Figure \ref{fig.examples}. Further growth is not observed for a longer duration.

Samples are always measured at about 900\,K heater temperature.
Therefore, even if the tempering took place at higher temperatures, the collisions were studied when the samples were cooler again. This way we do not trace any sticking properties related to viscosity effects, instead only due to compositional variations. With the given setup, we do not trace the phase transitions occurring at 700 or 800\,K as studied by \cite{deBeule2017}, rather we trace changes at higher temperatures.

We used two slightly different settings for the temperature of the heater, which result in different mean translation velocities ($v_{\mathrm{mean}}^{(1)} \approx 2$\,cm\,s$^{-1}$ for the first and $v_{\mathrm{mean}}^{(2)} \approx 1$\,cm\,s$^{-1}$ for the second heater setting) of the aggregates and thus also in different maximum aggregate sizes. The difference is due to the largest aggregates encountering other aggregates at velocities beyond their fragmentation limit ($v_{\mathrm{max}}\approx 0.2$\,m\,s$^{-1}$). This reduces the absolute size and shows that details of the setting are important. However, by leaving all of parameters the same for a certain study, the influence of tempering can be traced by the maximum aggregate size. Here, the important findings are relative changes of aggregate size with the temperature used for heating the aggregates. 

\section{Samples}

The idea behind these experiments is that a dust sample that drifts inward within a 
protoplanetary disk is subject to increasing temperatures, which changes the composition and, related to this, the grain size and sticking properties. In particular for dominating species like pyroxenes and olivines, phase transitions already occur at moderate temperatures of several hundred K, which might be relevant \citep{deBeule2017}. How this really affects collisional growth has not been studied before.
The basic requirement to test our hypothesis is therefore a sample with silicates, and 
for convenience we chose a basalt sample here.

\subsection{Grain sizes}
\begin{figure}[h]
	\includegraphics[width=\columnwidth]{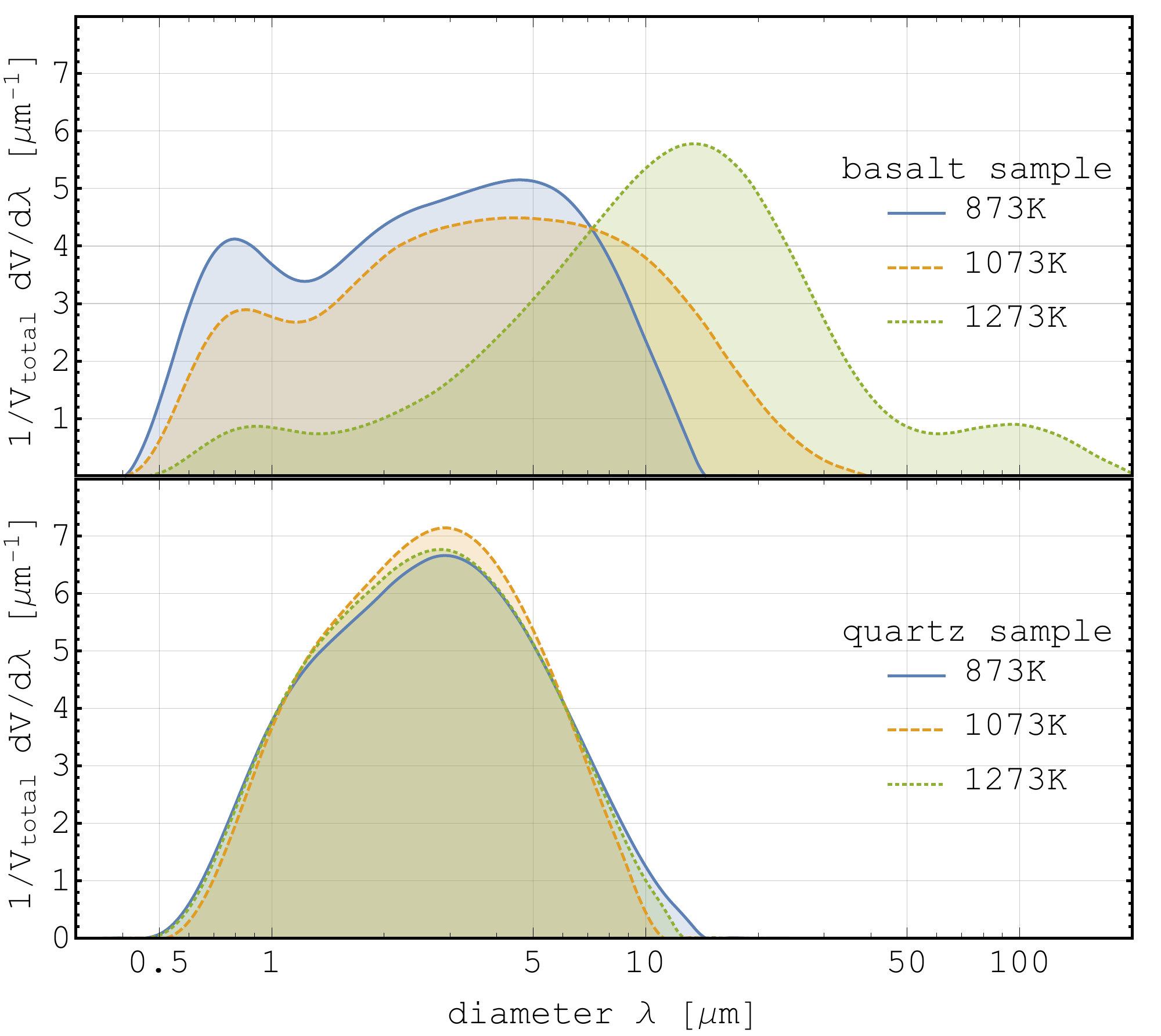}
	\caption{\label{fig.grainsizevolume}Volume size distributions of dust grains in the different samples; the distributions were measured by a commercial instrument based on light scattering (Malvern Mastersizer 3000).}
\end{figure}

\begin{figure}[h]
	\includegraphics[width=\columnwidth]{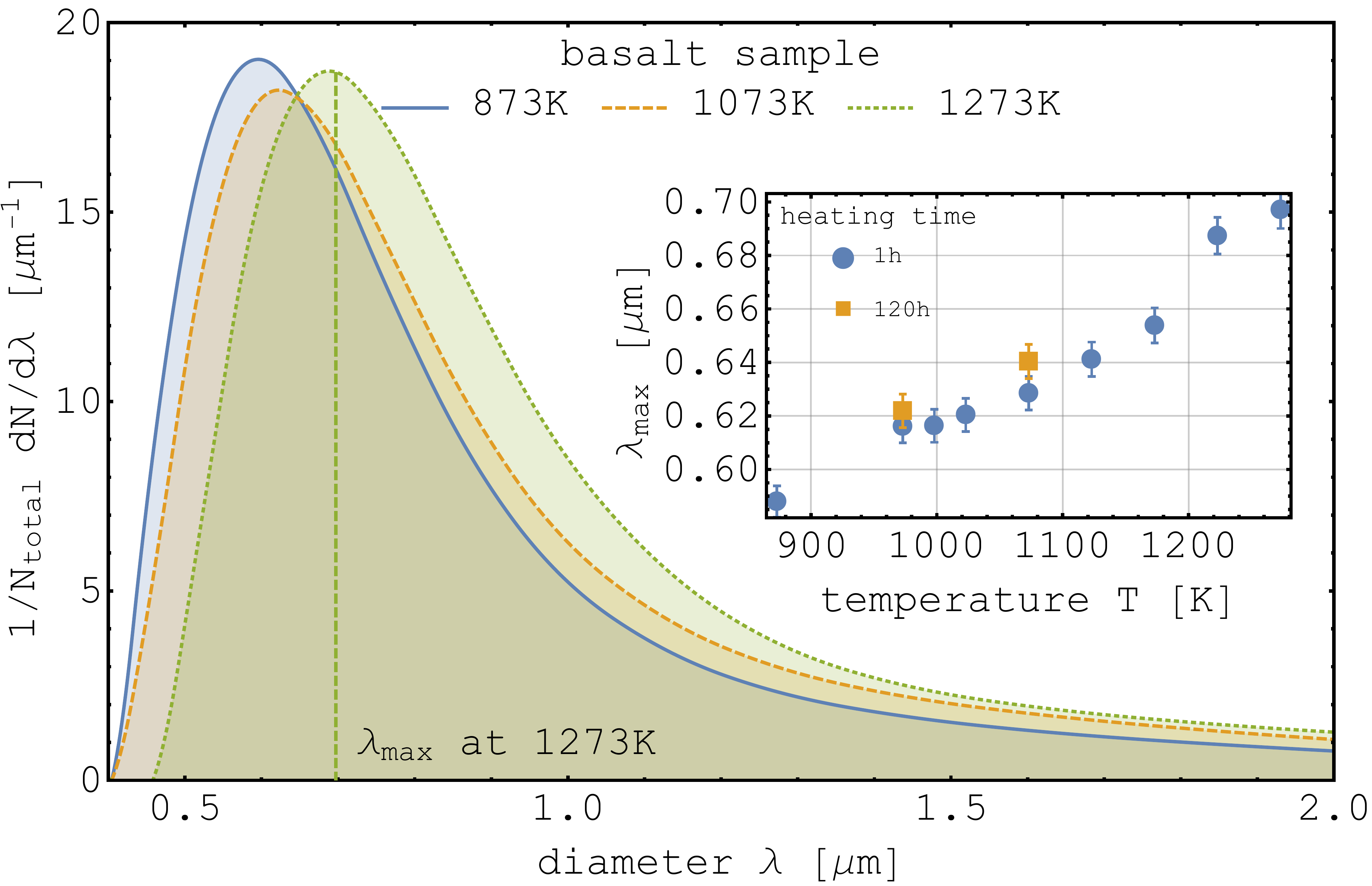}
	\caption{\label{fig.grainsizenumber}Number size distributions of dust grains in the different samples (Malvern Mastersizer 3000). The inset shows the most likely dust grain diameter $\lambda_{\mathrm{max}}$ for the different heating temperatures and heating times.}
\end{figure}

The basic sample is produced by milling a basalt sample to grain sizes in the micrometer range as expected 
for dust in protoplanetary disks.
From this sample parts are used for tempering at different temperatures, and from each subsample a part was used for detailed grain size analysis by a 
particle sizer (Mastersizer 3000).
For comparison we also used a quartz sample where we would not expect any changes upon tempering and in collision experiments. 
The volume size distribution puts emphasis on the grains dominating in mass and examples are shown in Figure \ref{fig.grainsizevolume}. Number size distributions are shown in Figure \ref{fig.grainsizenumber}.

The number-dominating as well as the volume-dominating grain sizes continuously increase with temperature for the basalt samples. Possible causes for this include compositional changes (see below). Sintering could also be a reason for the observed increase in the grain sizes. The 120\,h basalt sample heated at a temperature below 1000\,K shows no significant change compared to the 1\,h sample.
For the sample heated above 1000\,K the grain size is somewhat larger compared to the 1 hour sample. This indicates that time has an influence here. However, this would only support our general findings on collisional evolution as given below.
As expected, we can see no significant changes for the pure quartz samples. Tab. \ref{tab:grainsize} summarizes the most likely grain sizes $\lambda_{\mathrm{max}}$ and the mean grain sizes $\lambda_{\mathrm{mean}}$ in dependence of the tempering temperature and tempering duration.

\startlongtable
\begin{deluxetable}{ccccc}
	\tablecaption{\label{tab:grainsize} Most Likely Grain Size $\lambda_{\mathrm{max}}$ and Mean Grain Size $\lambda_{\mathrm{mean}}$ of the Used Samples in Dependence of the Tempering Temperature and Tempering Duration.}
	\tablehead{
		\colhead{Material} & \colhead{Temperature} & \colhead{Duration} & \colhead{$\lambda_{\mathrm{max}}$\tablenotemark{a}} & \colhead{$\lambda_{\mathrm{mean}}$\tablenotemark{b}} \\
	& 	\colhead{(K)} & \colhead{(hour)} & \colhead{($\mu$m)} & \colhead{($\mu$m)}
	}
	\startdata
	basalt & 873 & 1 & 0.588 &$3.5\pm0.3$\\
	basalt & 973 & 1 & 0.616 &$6.1\pm0.6$\\
	basalt & 998 & 1 & 0.616 &$3.4\pm0.3$\\
	basalt & 1023 & 1  & 0.620 &$4.2\pm0.5$\\
	basalt & 1073 & 1  & 0.629 &$5.6\pm0.7$\\
	basalt & 1123 & 1  & 0.641 &$5.2\pm0.5$\\
	basalt & 1173 & 1  & 0.654&$8.2\pm0.9$\\
	basalt & 1223 & 1  & 0.687&$10.7\pm1.2$\\
	basalt & 1273 & 1  & 0.697&$18.9\pm4.8$\\
	basalt & 973 & 120  & 0.622&$5.7\pm0.6$\\
	basalt & 1073 & 120 & 0.64&$5.1\pm0.5$\\
	quartz & 873 & 1 & - &$3.3\pm0.3$\\
	quartz & 1073 & 1 & - &$3.1\pm0.2$\\
	quartz & 1273 & 1 & - &$3.2\pm0.2$\\
	\enddata
	\tablenotetext{a}{Determined from the number size\\
		distributions (see Figure \ref{fig.grainsizenumber}). Tolerance $\Delta \lambda_{\mathrm{max}}=0.006\,\mu$m.}
	\tablenotetext{b}{Determined from the volume size distributions (see Figure \ref{fig.grainsizevolume})}
\end{deluxetable}

\subsection{Composition}
\label{sec:composition}

To investigate possible changes in sample composition, M{\"o}ssbauer spectra of basaltic dust heated to 873-1073\,K for 1\,h were recorded in transmission geometry and constant acceleration mode at 80\,K using a l-He bath cryostat.

\begin{figure}[h]
	\includegraphics[width=\columnwidth]{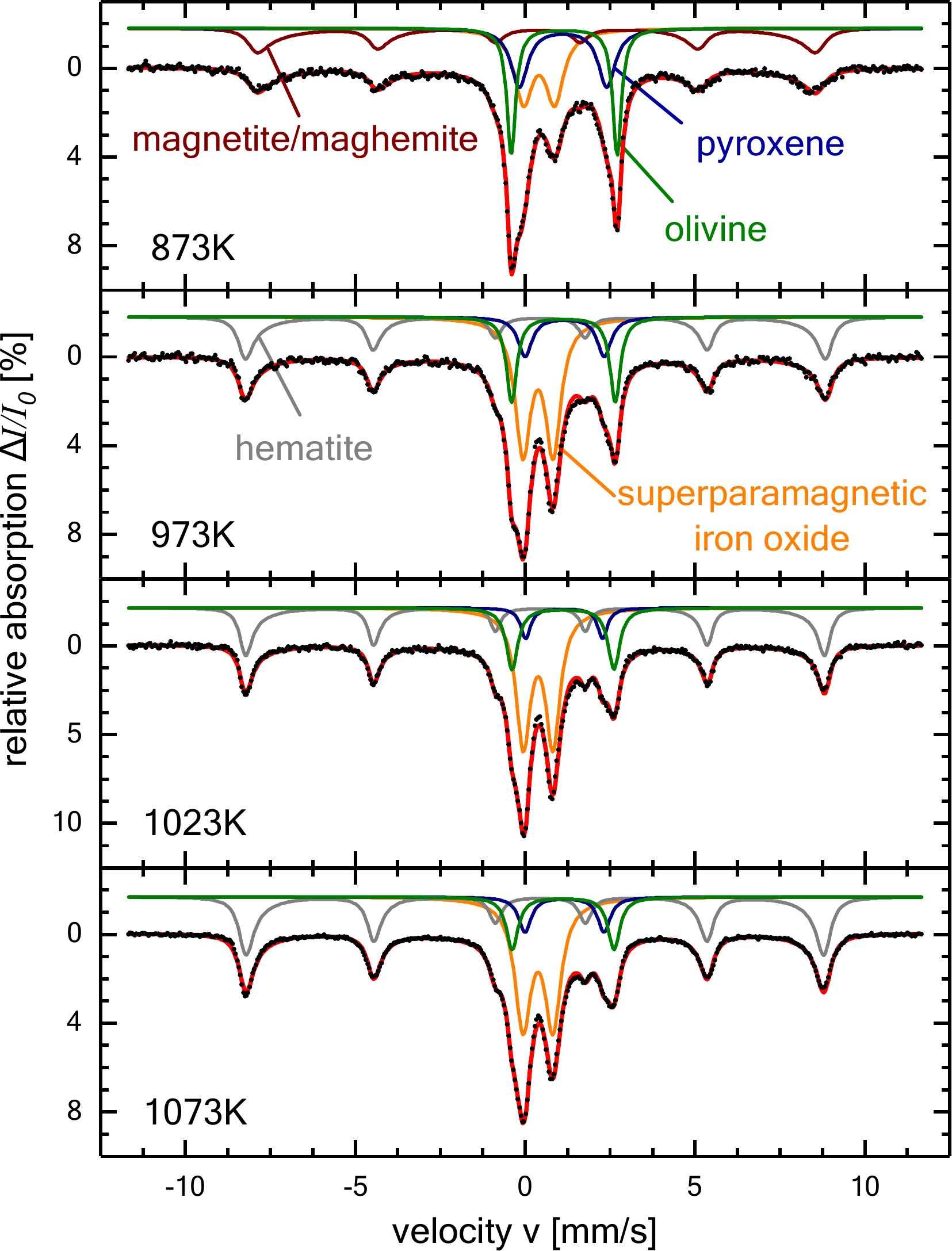}
	\caption{\label{fig.mossbauer}M{\"o}ssbauer spectra of basaltic dust heated at 873-1073\,K recorded at 80\,K.}
\end{figure}

The spectrum of material heated to 873\,K shown in the top panel of Figure \ref{fig.mossbauer} displays a complex spectral structure containing several subspectra. Two dominant doublet subspectra with high isomer shift $\delta$ and quadrupole splitting $\Delta E_Q$ are visible, indicating Fe$^{2+}$-bearing paramagnetic minerals. Absolute numbers of $\delta \approx 1.25$\,mm\,s$^{-1}$ (relative to $\alpha$-Fe at ambient temperature) and $\Delta E_Q$ of ca. $3.0$\,mm\,s$^{-1}$ and $2.3$\,mm\,s$^{-1}$, respectively, are consistent with literature values reported for iron-bearing minerals of the olivine (green) and pyroxene (blue) group \citep{Oshtrakh2007}. Additionally, the spectrum contains a minor doublet of smaller quadrupole splitting and isomer shift (orange) and a broad asymmetric sextet (dark red), whose hyperfine parameters point toward ferric oxide. This combination implies the presence of nanophase ferric oxide with partial superparamagnetic properties, although the presence of additional ferric paramagnetic minerals cannot be ruled out \citep{Morris1993}.

\begin{figure}[h]
	\includegraphics[width=\columnwidth]{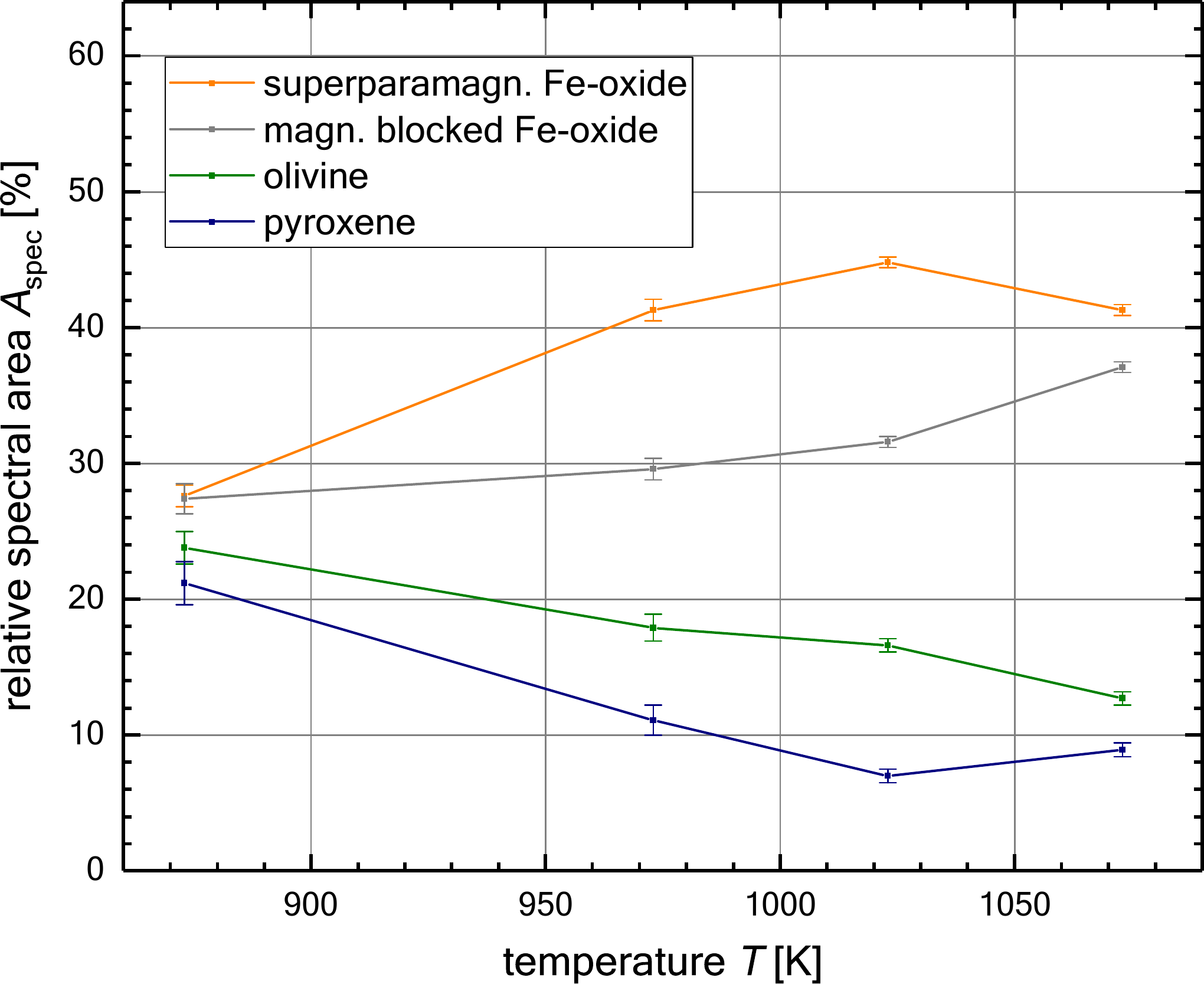}
	\caption{\label{fig.spectralArea}Relative spectral areas of identified mineral subspectra in M{\"o}ssbauer spectra of heated basaltic dust.}
\end{figure}

Spectra of basaltic dust heated to higher temperatures display widely similar spectral structures. However, the spectral areas of olivine and pyroxene decrease upon heating to the benefit of superparamagnetic as well as magnetically blocked ferric oxide. Relative spectral areas of the identified mineral subspectra are displayed comparatively in Figure \ref{fig.spectralArea} for different heating temperatures to allow a better understanding of phase transitions and possible structural changes of the basaltic material upon heating. Presumably, the decrease in olivine and pyroxene spectral area corresponds to a partial phase transition of Fe$^{2+}$-bearing silicates to iron oxide, which takes place at temperatures up to about 1000\,K. A partial phase transition refers to the fact that the transition does not occur for the entire material.

Upon further inspection, a continuous decrease in sextet line width is evident, as well as a change of the sextet quadrupole level shift from  $\sim0$\,mm\,s$^{-1}$ to $-0.2$\,mm\,s$^{-1}$. This value is characteristic for hematite above the Morin transition, which is known to be suppressed in hematite nanocrystals \citep{Ozdemir2008}. Thereby, it may indicate a phase transition from ferrimagnetic magnetite and/or maghemite (dark red) to antiferromagnetic hematite nanocrystals (gray). The decrease in line width, on the other hand, could originate from structural ordering or, alternatively, from a moderate increase in hematite crystal size, resulting in a minor variation of superparamagnetic relaxation.

\section{Bouncing Barrier Size}

\begin{figure}[h]
	\includegraphics[width=\columnwidth]{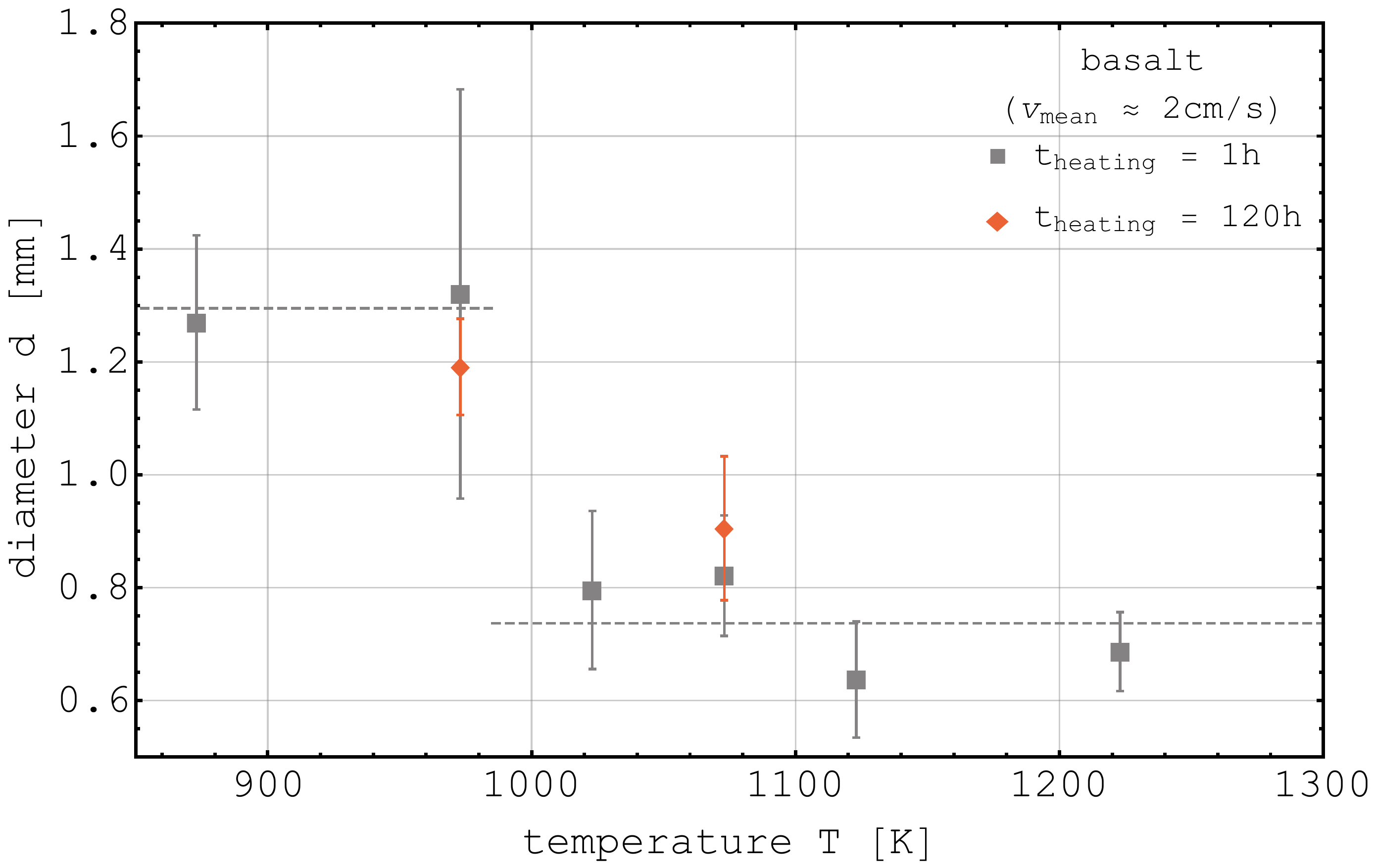}
	\caption{\label{fig.result1}Maximum size of aggregates grown for samples tempered at different temperatures for heater setting 1. The dashed lines mark the average values below and above the 1000\,K value that seems to divide two regimes. The average value below $1000$\,K is $1.75 \pm 0.16$ times the value above $1000$\,K.
	The basalt samples heated for 120\,h do not differ within the accuracy from the 1\,h samples. The displayed data points are average values of 49 measurements.}
\end{figure}

\begin{figure}[h]
	\includegraphics[width=\columnwidth]{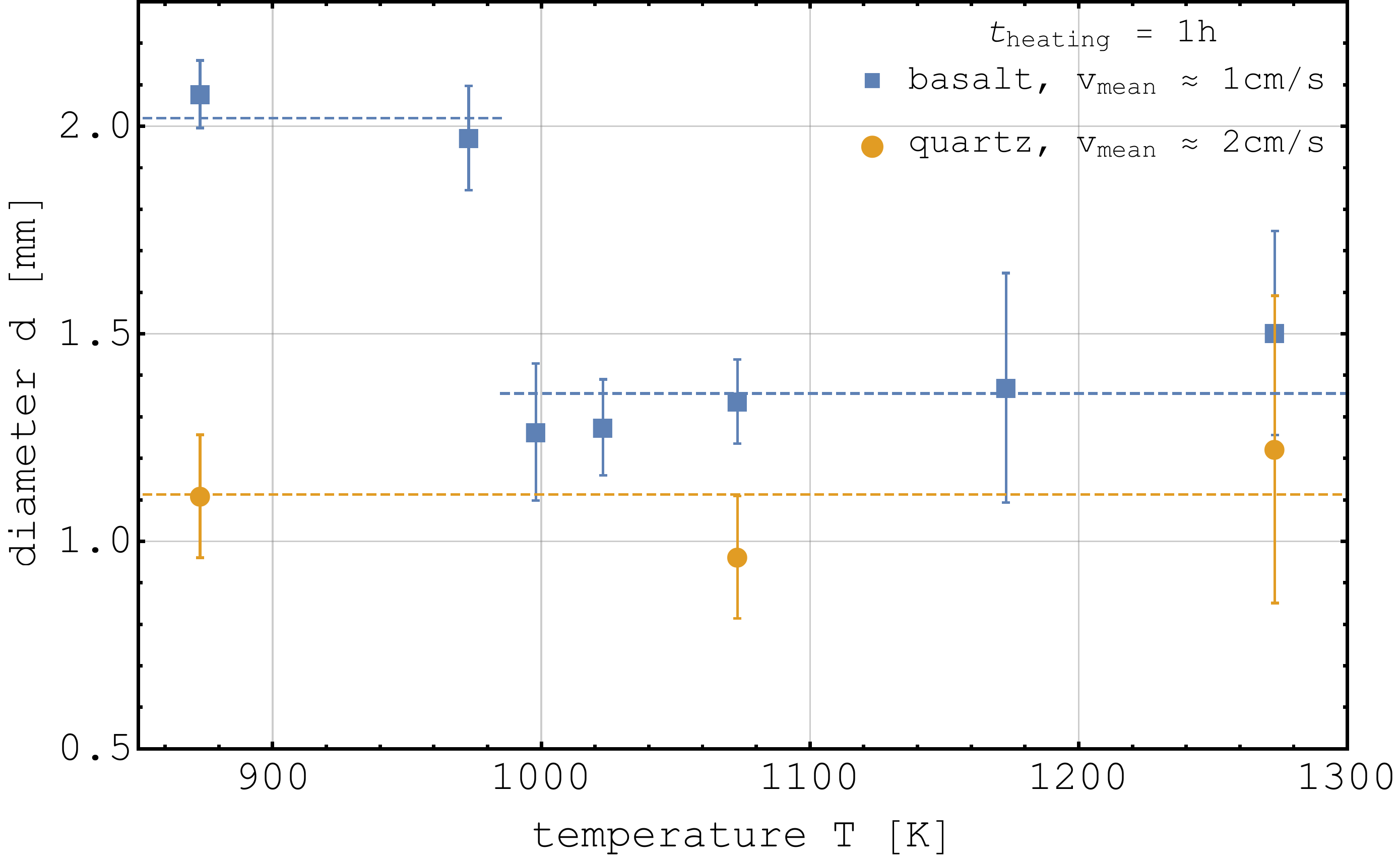}
	\caption{\label{fig.result2}Same as Figure \ref{fig.result1}, but for different heater temperature (basalt sample). The average value below $1000$\,K is $1.49 \pm 0.08$ times the value above $1000$\,K. In addition, results from the quartz samples are shown. The displayed data points are average values of 69 measurements.}
\end{figure}

Figure \ref{fig.result1} and \ref{fig.result2} show the maximum aggregate sizes that formed for the different dust samples tempered at different temperatures. The maximum aggregate size $d=2\,\sqrt{\frac{A}{\pi}}$ is determined by the area $A$ of the two-dimensional projection of the aggregate.

There is a clear offset in growth at $1000$\,K. Dust tempered at lower temperatures grows larger than the dust processed at higher temperatures. This is true for both heater settings. The ratio of the average diameter below $1000$\,K to the value above this temperature is $1.75 \pm 0.16$ for $v_{\mathrm{mean}}^{(1)} \approx 2$\,cm\,s$^{-1}$. For $v_{\mathrm{mean}}^{(2)} \approx 1$\,cm\,s$^{-1}$ this size ratio is $1.49 \pm 0.08$.
For comparison we also measured the largest aggregate size for a pure quartz dust sample (see Figure \ref{fig.result2}).
These quartz results confirm that without changes in composition we do not see any change in growth. It also implies that the jump in aggregate size is obviously not just an artifact by the thermal processing and sample handling.

It is highly probable that internal phase transitions are responsible for the change in growth.
Grain size increase might be a reason for less sticking.
However, the number-dominating size does not change spontaneously across the transition, rather it increases continuously.
We also do not see a sudden transition at temperatures near 1000\,K in the mean grain size for the volume-dominating grains. Even the M{\"o}ssbauer spectra do not show a sudden change near this temperature.
We therefore cannot pin down one special reason why sticking changes at a certain
temperature. \cite{deBeule2017} saw strong morphological changes on the grains of
their samples at lower temperatures. Whether grain morphology is also important for the smaller grains studied here is up to future work. The M{\"o}ssbauer spectra may indicate a phase transition and/or a structural ordering (see sec. \ref{sec:composition}). This could change the grain surface, but without further investigation, we can neither confirm nor exclude this presumption.
This does not devalue the present study though. It is a fact that the growth
changes with temperature whatever the specific reason.

\section{Conclusion}

Our results show that aggregate growth in protoplanetary disks might depend on the temperature history.
A certain (large) size might be a prerequisite for efficient concentration mechanisms like
the streaming instability \citep{Youdin2005,Johansen2006}.
If so, larger aggregate growth might promote the local formation of planetesimals ''outside of 1000\,K''.
Certainly, modifications of orbits occur at later times but from a probabilistic point
of view there might be some memory left from the planetesimal formation process
and this indeed seems to correlate to the exoplanet database. 

We see a strong correlation between the temperature of a dust sample and the
maximum size of aggregates formed until they reach the bouncing barrier.
There is a clear separation in favored growth below 1000\,K and suppressed growth
beyond 1000\,K. The same correlation is found between local temperature and the occurrence of extrasolar terrestrial planets, which seem to be sparse at temperatures higher than 1000\,K.
There are still some steps in between millimeter dust and planets. However, planetesimal formation, e.g., triggered by streaming instabilities, would favor larger aggregates, and terrestrial planets might form locally. Assuming this, the almost too-perfect
correlation between early dust growth and planet locations is tantalizing. 
As a caveat, we only considered a special dust sample in the experiments and, e.g. a 
heater in air, so this is certainly still far from optimized. Also, the local temperatures might vary somewhat at the times planetesimals formed.
Taken together, future experiments need to show whether 1000\,K is a general mark
showing a transition in collision experiments for a wider range of mineral compositions of dust. However, our results already strongly indicate that the growth of millimeter-size dust at high temperatures might be a key factor in the formation of
terrestrial planets in the inner regions of protoplanetary disks.

\section*{Acknowledgements}
This project is funded by DFG grant KE 1897/1-1, SPP 1681 grant WE2623/7-2, and FOR 1509 grant WE2623/13. We also thank the referee for a constructive review of the paper.




\begin{thebibliography}{}
	
\bibitem[Alibert et al., 2005]{Alibert2005}
Alibert, Y., Mordasini, C., Benz, W., \& Winisdoerffer, C. 2005, A\&A, 434, 343	
	
\bibitem[Aumatell \& Wurm, 2013]{Aumatell2013}
Aumatell, G., \& Wurm, G. 2013, MNRAS, 437, 690

\bibitem[Benz et al., 2014]{Benz2014}
Benz, W., Ida, S., Alibert,  Y., Lin, D., \& Mordasini, C. 2014, in Protostars \& Planets VI, ed. H. Beuther et al. (Tucson, AZ: Univ. Arizona Press), 691

\bibitem[Blum \& Wurm, 2008]{Blum2008}
Blum, J., \& Wurm, G. 2008, ARA\&A, 46, 21

\bibitem[Boley \& Ford, 2013]{Boley2013}
Boley, A.~C., \& Ford, E.~B. 2013, arXiv, 1306.0566

\bibitem[Chiang \& Youdin, 2010]{Chiang2010}
Chiang, E., \& Youdin, A.~N. 2010, AREPS, 38, 493

\bibitem[de Beule et al., 2015]{deBeule2015}
de Beule, C., Wurm, G., Kelling, T., Koester, M., \& Kocifaj, M. 2015, Icar, 260, 23

\bibitem[de Beule et al., 2017]{deBeule2017}
de Beule, C., Landers, J., Salamon, S., Wende, H., \& Wurm, G. 2017, ApJ, 837, 59

\bibitem[Dominik \& Tielens, 1997]{Dominik1997}
Dominik, C., \& Tielens, A.~G.~G.~M. 1997, ApJ, 480, 647

\bibitem[Dr{\c a}{\.z}kowska \& Dullemond, 2014]{Drazkowska2014}
Dr{\c a}{\.z}kowska, J., \& Dullemond, C.~P. 2014, A\&A, 572, A78

\bibitem[D{\"u}rmann \& Kley, 2015]{Duermann2015}
D{\"u}rmann, C., \& Kley, W. 2015,  A\&A, 574, A52

\bibitem[D{\"u}rmann \& Kley, 2017]{Duermann2017}
D{\"u}rmann, C., \& Kley, W. 2017,  A\&A, 598, A80

\bibitem[Ebel \& Alexander, 2011]{Ebel2011}
Ebel, D.~S., \& Alexander, C.~M.~O'~D. 2011, P\&SS, 59, 1888

\bibitem[exoplanet.eu, 2017]{Exoplanet.eu}
exoplanet.eu, ''The Extrasolar Planets Encyclopedia'', 2017 July 10 (www.exoplanet.eu)

\bibitem[Gundlach et al., 2011]{Gundlach2011}
Gundlach, B., Kilias, S., Beitz, E., \& Blum, J. 2011, Icar, 214, 717

\bibitem[Gundlach \& Blum, 2015]{Gundlach2015}
Gundlach, B., \& Blum, J. 2015, ApJ, 798, 34

\bibitem[G\"uttler et al., 2010]{Guettler2010}
G\"uttler, C., Blum, J., Zsom, A., Ormel, C. W., \& Dullemond, C.~P. 2010, A\&A,
513, A56

\bibitem[Hayashi et al., 1985]{Hayashi1985}
Hayashi, C., Nakazawa, K., \& Nakagawa, Y. 1985, Protostars and Planets II
(Tucson, AZ: Univ. Arizona Press), 1100

\bibitem[Hu et al., 2014]{Hu2014}
Hu, X., Tan, J.~C, \& Chatterjee, S. 2014, in IAU Symp. 310, Complex Planetary Systems, ed. Z. Kne\v{z}evi\'{c} \& A. Lema\^{i}tre (Cambridge: Cambridge Univ. Press), 66

\bibitem[Jankowski et al., 2012]{Jankowski2012}
Jankowski, T., Wurm, G., Kelling, T., Teiser, J., Sabolo, W., Guti{\'e}rrez, P.~J., \& Bertini, I. 2012, A\&A, 542, 80

\bibitem[Johansen et al., 2006]{Johansen2006}
Johansen, A., Klahr, H., \& Henning, Th. 2006, ApJ, 636, 1134

\bibitem[Kelling et al., 2014]{Kelling2014}
Kelling, T., Wurm, G., \& Koester, M. 2014, ApJ, 783, 111

\bibitem[Klahr \& Henning, 1997]{Klahr1997}
Klahr, H., \& Henning, T. 1997, Icar, 128, 213

\bibitem[Kley, 2000]{Kley2000}
Kley, W. 2000, MNRAS, 313, L47

\bibitem[Knudsen, 1909]{Knudsen1909}
Knudsen, M. 1909, AnPhy, 336, 633

\bibitem[Koester et al., 2017]{Koester2017}
Koester, M., Kelling, T., Teiser, J., \& Wurm, G. 2017, Ap\&SS, 362, 171

\bibitem[Kruss et al., 2016]{Kruss2016}
Kruss, M., Demirci, T., Koester, M., Kelling, T., \& Wurm, G. 2016, ApJ, 827,
110

\bibitem[Kruss et al., 2017]{Kruss2017}
Kruss, M., Teiser, J., \& Wurm, G. 2017, A\&A, 600, A103

\bibitem[Morris et al., 1993]{Morris1993}
Morris, R.~V., Golden, D.~C., Bell III, H.~V. Lauer Jr., R.~B. \& Adams, J.~B. 1993 GeCoA, 57, 4597

\bibitem[Muntz et al., 2002]{Muntz2002}
Muntz, E. P., Sone, Y., Aoki, K., Vargo, S., \& Young, M. 2002, JVSTA, 20, 214

\bibitem[Musiolik et al., 2016a]{Musiolik1}
Musiolik, G., Teiser, J., Jankowski, T., \& Wurm, G. 2016, ApJ, 818, 16

\bibitem[Musiolik et al., 2016b]{Musiolik2}
Musiolik, G., Teiser, J., Jankowski, T., \& Wurm, G. 2016, ApJ, 827, 63

\bibitem[Nelson \& Benz, 2003]{Nelson2003}
Nelson, A.~F., \& Benz, W. 2003, ApJ, 589, 556

\bibitem[Okuzumi et al., 2012]{Okuzumi2012}
Okuzumi, S., Tanaka, H., Kobayashi, H., \& Wada, K. 2012, ApJ, 752, 106

\bibitem[Oshtrakh et al., 2007]{Oshtrakh2007}
Oshtrakh, M.~I., Petrova, E.~V., Grokhovsky, V.~I., \& Semionkin, V.~A. 2007, HyInt 177, 65

\bibitem[{\"O}zdemir et al., 2008]{Ozdemir2008}
{\"O}zdemir, {\"O}, Dunlop, D.~C., \& Berqu{\'o}, T.~S. 2008, GGG, 9, Q10Z01

\bibitem[Schneider et al., 2012]{Schneider2012}
Schneider, J., Le Sidaner, P., Savalle, R., \& Zolotukhin, I. 2012, adass XXI, 461, 447

\bibitem[Scott, 2007]{Scott2007}
Scott, E.~R.~D., 2007, AREPS, 35, 577

\bibitem[Steinpilz et al., 2017]{Steinpilz2017}
Steinpilz, T., Teiser, J., Koester, M., Schywek, M., \& Wurm, G. 2017, MiST, 29, 235

\bibitem[Wada et al., 2011]{Wada2011}
Wada, K., Tanaka, H., Suyama, T., Kimura, H., \& Yamamoto, T. 2011, ApJ, 737, 36

\bibitem[Weidenschilling, 1977]{Weidenschilling1977}
Weidenschilling, S.~J. 1977, MNRAS, 180, 57

\bibitem[Wurm et al., 2013]{Wurm2013}
Wurm, G., Trieloff, M., \& Rauer, H. 2013, ApJ, 769, 78

\bibitem[Yang et al., 2017]{Yang2017}
Yang, C.~C., Johansen, A., \& Carrera, D., 2017, A\&A, in press (arXiv:1611.
07014)

\bibitem[Youdin \& Goodman, 2005]{Youdin2005}
Youdin, A.~N., \& Goodman, J. 2005, ApJ, 620, 459

\bibitem[Zsom et al., 2010]{Zsom2010}
Zsom, A., Ormel, C.W., G{\"u}ttler, C., Blum, J., \& Dullemond, C.~P. 2010, A\&A, 513, A57 

\end{thebibliography}
\end{document}